\documentclass[seqeq,twocolumn]{jpsj3}
\def\D{{\rm d}}
\def\I{{\rm i}}
\def\E{{\rm e}}
\def\be{\begin{eqnarray}}
\def\ee{\end{eqnarray}}
\def\beq{\begin{eqnarray}}
\def\eeq{\end{eqnarray}}
\title{Stability of the Excitonic Phase in Bilayer Quantum Hall Systems\\
 at Total Filling One\\
-- Effects of Finite Well Width and Pseudopotentials --}

\author{Daijiro {\sc Yoshioka}\thanks{E-mail address: daijiro@toki.c.u-tokyo.ac.jp} and $^{1}$Naokazu {\sc Shibata}} 

\inst{Department of Basic Science, University of Tokyo, 3-8-1 Komaba, Meguro-ku, Tokyo 153-8902 \\
$^{1}$Department of Physics, Tohoku University, 
Aoba, Aoba-ku, Sendai 980-8578
}
\abst{
The ground state of a bilayer quantum Hall system at $\nu_{\rm T}=1$ with model pseudopotential is investigated by the DMRG method.
Firstly, pseudopotential parameters appropriate for the system with finite layer thickness are derived, and it is found that the finite thickness makes the excitonic phase more stable.
Secondly, a model, where only a few pseudopotentials with small relative angular momentum have finite values, is studied, and it is clarified how the excitonic phase is destroyed as intra-layer pseudopotential becomes larger.
The importance of the intra-layer repulsive interaction at distance twice of the magnetic length for the destruction of the excitonic phase is found.
}

\kword{bilayer quantum Hall system, excitonic state, phase transition, composite fermion, strong magnetic field, DMRG}

\begin{document}
\maketitle

\section{Introduction}

The bilayer quantum Hall system at total filling factor $\nu_{\rm T}=1$ is known to be in the inter-layer coherent state when the layer separation is small enough.\cite{YMG,YM,WZ,EI,MMY}
The coherent state in the limit of vanishing layer separation is described by Halperin's $\Psi_{111}$ state\cite{Halperin}, which can also be considered as an excitonic state, where an electron in one of the layer is bound to a hole in the other layer.
Many experimental evidences for this excitonic correlation have been accumulated.\cite{ZTNC,ZTNC1,CFH0,CFH1,CFH2}
The excitonic correlation becomes gradually smaller as the layer separation increases.
It has been clarified that the excitonic correlation almost vanishes at around $(d/l)_{\rm c} = 1.85$ for typical samples,\cite{ZTNC,ZTNC1,CFH0,CFH1} where $d$ is the distance between the two layers and $l=\sqrt{\hbar/eB}$ is the magnetic length.\cite{note1}
At larger layer separation, the two layers become independent of the other layer, and each layer is in the composite fermion (CF) liquid state.\cite{Jain,Jain1,Halp}

The details of the transition between the excitonic phase and the independent CF liquid state, however, have not been clarified in spite of intense theoretical\cite{HFT,Zheng,Narash,PCF,Mac,Kim,Burkov,SH,NY,SRM,Shib2,Moller} and experimental investigations.\cite{Wies,ZTNC2,Kumada,CHAMP,Muraki,Finck,Muraki1}
There are several issues to be clarified both theoretically and experimentally.
The first issue is what is the critical value of $d/l$ at which the excitonic phase is terminated.
The second issue is whether the transition is continuous or discontinuous.
The third issue is whether electron spin is involved in this transition or not.
Actually, these issues are related to each other.

As for the involvement of the spin degree of freedom, all the theoretical investigations have assumed full spin polarization both in the CF liquid phase and in the excitonic phase.
This is because the system becomes much simpler if we can assume full spin polarization, and we can expect
the spin degree of freedom can be suppressed in the strong magnetic field where Zeeman splitting is large.
Furthermore, full spin polarization in the excitonic phase is expected even in smaller Zeeman splitting from the following theoretical consideration.
Namely, it should be noticed that the excitonic state can also be understood as pseudospin ferromagnetic state, where the $z$ component of the pseudospin stands for the layer index.\cite{YMG}
The Coulomb interaction between electrons in the different layers forces the pseudospin align in the $xy$-plane, and the excitonic state is realized.
Considering the facts that the interaction between electrons with opposite spin is stronger than that between electrons in different layers at finite $d/l$, and the existence of the Zeeman splitting, real spin of the electron must also be ferromagnetically aligned when the excitonic state (i.e. pseudospin ferromagnetic state) is realized.
Thus, the theoretical estimates for the boundary value $(d/l)_{\rm c}$ have been done between spin-polarized excitonic phase and the spin-polarized CF liquid phase.
This assumption also leaves a possibility for continuous transition between the two phases.

With this assumption of the full spin polarization, the estimates of $(d/l)_{\rm c}$ is not so different from theory to theory.
However, even in this case, there has been no consensus on the details of the transition.
Some of the theories suggest a first order phase transition,\cite{Mac,Burkov,SH} and the others suggest second order phase transition.\cite{Shib2}
The possibility of smooth crossover between the two limiting states has not been denied.\cite{SRM,Moller}
Possibility of intermediate states is also raised.\cite{HFT,Zheng,Narash}

The situation become quite different if the CF liquid phase is partially spin-polarized.
The transition between the two states with different spin polarization cannot be continuous.
Moreover, the estimate of $d/l$ should be reconsidered.
This is because the partial spin polarization in the CF liquid phase means that the spin-polarized CF liquid has higher energy, so the boundary between the excitonic phase should have moved to the lower value of $d/l$ than the estimate assuming full polarization.

Now, recent experiments are giving evidences that the spin degree of freedom is important.\cite{ZTNC2,Kumada,Muraki,Finck}
It has been found that there is a first order phase transition between excitonic state and partially spin-polarized CF liquid state in typical experimental situations, although the transition looks like continuous possibly due to inhomogeneity of the sample.
This fact has raised three questions: (i) what property the state in the partial spin polarization has? (ii) how the transition occurs if the Zeeman splitting is enhanced? and (iii) what is the value of $(d/l)_{\rm c}$, if enhanced Zeeman splitting makes both phase spin-polarized?
For the first question theoretical investigation has not been developed, mainly because the nature of the partially spin-polarized CF liquid state has not been clarified.
For the second and third questions, it has been found experimentally that the excitonic phase persists to larger layer separation, when the spin Zeeman splitting is enhanced by in-plane magnetic field, and the transition between polarized states now becomes continuous.\cite{Muraki,Finck}
In one of the experiments, value as large as $(d/l)_{\rm c} \simeq 2.3$ is reported.\cite{Muraki}

The large value of the transition point for spin-polarized system is surprising, because theoretical estimate for the transition point is around $d/l=1.8$ or smaller.\cite{MMY,HFT,Mac,Shib2,Moller}
We previously have investigated fully spin-polarized system by DMRG method.\cite{Shib2}
In that paper we considered a system where the thickness of each layer is negligibly small.
We calculated the excitonic correlation, and found that it decreases as the layer separation becomes larger, and almost vanishes at around $d/l=1.6$.

In the present paper, we restrict ourselves to fully spin-polarized systems as before.
We try to understand the shift of the phase boundary to larger value of $d/l$, and try to understand phase transition in such spin-polarized systems.
We suspect this discrepancy of the transition point between theory and experiment comes from the thickness of the two-dimensional layer.
Thus, in section 2 we calculate pseudopotential parameters taking into account thickness effect, and using those parameters we determine the transition point where the excitonic correlation vanishes.
In order to investigate the mechanism of the transition in the spin-polarized system further, we consider in section 3 a minimal model in which only a few pseudopotential parameters are included.
Discussion and summary of the results are given in section 4.
Investigation of the partially spin-polarized CF liquid state is left for the future work.

\section{Effect of Layer Thickness}

\subsection{Pseudopotentials}

Previously, development of the ground state as the layer separation increases has been mostly studied neglecting the thickness of the quantum wells making up the bilayer system.\cite{note}
In this paper we calculate Haldane's pseudopotentials\cite{Haldane} considering the layer thickness, and investigate the effect on the ground state.

We consider a model bilayer system where each layer is realized as a quantum well with infinitely high confining potential.
The width of the well is denoted as $w$, and the separation between the layers $d$ is measured between the center of each well.
We take the $z$-axis perpendicular to the 2-d layers, so that the electrons can reside either at $-(d+w)/2 \le z \le -(d-w)/2$, or at $(d-w)/2 \le z \le (d+w)/2$ in this model.
The single particle wave function is written as
\begin{equation}
\Psi_{\pm}(x,y,z) = \phi(x,y)\psi_{\pm}(z),
\label{eq:1.1}
\end{equation}
where $+$ and $-$ distinguish the layers and
\begin{equation}
\psi_\pm(z)=\left\{
\begin{array}{lll}
\sqrt{\frac{2}{w}}\cos&\left(\frac{\pi (z \mp d/2)}{w}\right),
& \\
& \pm\frac{d}{2}-\frac{w}{2} \le z \le \pm\frac{d}{2}+\frac{w}{2}&\\[2mm]
0, & \text{otherwise.}&\\
\end{array}
\right.
\label{eq:1.2}
\end{equation}

Using these wave functions, we calculate Fourier transform of the inter-layer and intra-layer Coulomb interaction potentials, $V^{\rm e}(q)$ and $V^{\rm a}(q)$.
\begin{align}
V^i(q) = &\frac{e^2}{4\pi\epsilon q}\int\D z_1\int\D z_2 |\psi_\sigma(z_1)|^2
|\psi_{\sigma'}(z_2)|^2\E^{-|z_1-z_2|q},\,\nonumber\\
&(i=\text{a or e})\,,
\label{eq:1.3}
\end{align}
where $\epsilon$ is the dielectric constant, $\sigma$ and $\sigma'$ are either $+$ or $-$, $\sigma = \sigma'$ for intra-layer interaction $V^{\text{a}}(q)$, and $\sigma\ne\sigma'$ for inter-layer interaction
 $V^{\text{e}}(q)$
.

The integral can be done analytically.
The inter-layer interaction $V^\text{e}(q)$ is calculated as follows
\begin{align}
V^\text{e}(q) =& \frac{e^2}{4\pi\epsilon q} \,\left(\frac{2}{w}\right)^2 
\int_{-w/2}^{w/2} \D z_1 \int_{-w/2}^{w/2}\D z_2\cos^2\left(\frac{\pi z_1}{w} \right)\nonumber\\
& \times \cos^2\left(\frac{\pi z_2}{w} \right)\E^{-(z_1-z_2+d)q}\nonumber\\
=&
\frac{e^2}{4\pi\epsilon q}\left[\frac{4\pi^2}{wq(w^2q^2+4\pi^2)} \right]^2
\left(\E^{wq/2}-\E^{-wq/2} \right)^2\E^{-dq}\,.
\label{eq:1.5}
\end{align}

On the other hand, the intra-layer interaction is calculated as follows\cite{Price,note2}
\begin{align}
V^\text{a}(q) =& \frac{e^2}{4\pi\epsilon q} \,\left(\frac{2}{w}\right)^2 
\int_{-w/2}^{w/2} \D z_1 \int_{-w/2}^{w/2}\D z_2\cos^2\left(\frac{\pi z_1}{w} \right)\nonumber\\
&\times \cos^2\left(\frac{\pi z_2}{w} \right)\E^{-|z_1-z_2|q}\nonumber\\
=&
\frac{e^2}{4\pi\epsilon q}\frac{8}{(w^2q^2+4\pi^2)} 
\left[\frac{3}{8}wq + \frac{\pi^2}{wq} \right.\nonumber\\
& \left.- \frac{4\pi^4}{w^2q^2(w^2q^2+4\pi^2)}
\E^{-wq/2}\left(\E^{wq/2}-\E^{-wq/2}\right) \right]\,.
\label{eq:1.6}
\end{align}

Haldane's pseudopotential $V_m^i$ is obtained from $V^i(q)$:\cite{Haldane}
\be
V_m^i=\int_0^\infty \D q qV^i(q)L_m(q^2l^2)\E^{-q^2l^2},\, (i=\text{a or e})\,.
\label{eq:1.4}
\ee
This integral is evaluated numerically.
The results are plotted in Figs. 1 and 2 as relative deviation from the value at $w=0$ in percent as functions of $w/l$.
Namely, $\Delta V_m^\text{e}=(V_m^\text{e}/V_m^\text{e}(w=0)-1)\times 100$ for $m=0,1,2,3,4$ are plotted in Fig. 1, and  $\Delta V_m^\text{a}$ for $m=1,3,5$ are plotted in Fig.2.
We can see that the effect of finite $w$ is larger for smaller relative angular momentum $m$.
The inter-layer pseudopotentials $V_m^\text{e}$ increases for $m=0,1,2$, but other pseudopotentials decrease as $w$ increases.

\begin{figure}
\begin{center}
\includegraphics[width=8cm]{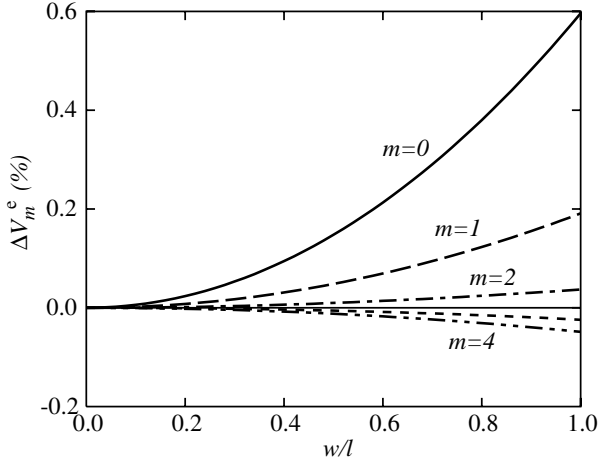}
\end{center}
\caption{
Relative deviation of the inter-layer pseudopotentials as functions of $w/l$ for a system with $d=2l$.
$\Delta V_m^\text{e}=(V_m^\text{e}/V_m^\text{e}(w=0)-1)\times 100$ are plotted as a solid line, dashed line, dash-dotted line, dotted-line, and dash-dot-dotted line for $m=0,1,2,3,4$, respectively.
}
\label{fig:1}
\end{figure}

\begin{figure}
\begin{center}
\includegraphics[width=8cm]{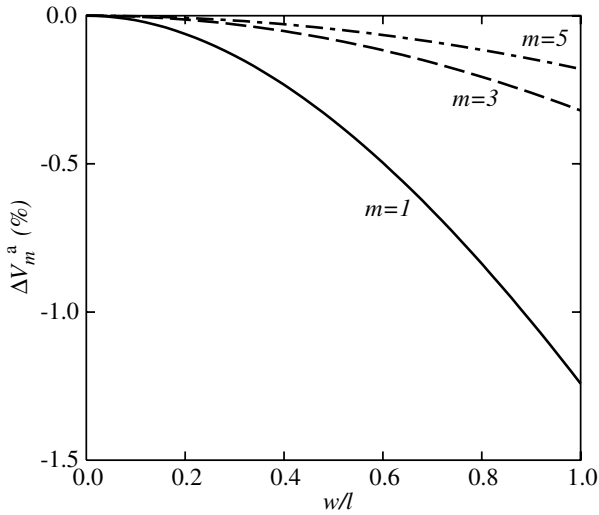}
\end{center}
\caption{
Relative deviation of the intra-layer pseudopotentials as functions of $w/l$.
$\Delta V_m^\text{a}=(V_m^\text{a}/V_m^\text{a}(w=0)-1)\times 100$ are plotted as a solid line, dashed line, and dash-dotted line for $m=1,3,5$, respectively.
}
\label{fig:2}
\end{figure}

\subsection{Results for the excitonic correlation}

We obtain the ground state of the bilayer system of finite electron number by DMRG method.\cite{DMRG,Shib,Shib1}
In this section we diagonalize rectangular systems with periodic boundary conditions, namely systems on torus, at total filling factor $\nu_{\rm T}=1$.
In order to see how the layer thickness affects the stability of the excitonic phase, we calculate excitonic correlation function for a system with finite $w$, and compare the results with our previous result obtained for a system with $w=0$.\cite{Shib2}
The excitonic correlation is defined by
\be
g_{\rm ex}(X)=\frac{N(N-1)}{N_1N_2}\langle \Psi|c_{\sigma,X}^\dagger c_{-\sigma,X} c_{-\sigma,0}^\dagger c_{\sigma,0}|\Psi\rangle,
\ee
where $|\Psi\rangle$ is the ground state and $c_{\sigma,X}^\dagger$ is the creation operator of the electron in the $\sigma$ layer at center coordinate $X$, $N$ is the total number of electrons, which is also the total number of single electron states per layer,  $N_1$ and $N_2$ are number of electrons in each layer ($N_1=N_2=N/2$).

In Fig. 3, we show results of the excitonic correlation $g_{\rm ex}(L_x/2)$ for systems with $w=0$ and $w/d=0.64$, where $L_x$ is the liner dimension of the system with periodic boundary condition.
The total number of electrons $N$ is $20$.
The value of $w/d$ is chosen to reproduce typical experimental situation where $d=28$\,nm and $w=18$\,nm.
As seen from the figure, the value of $d/l$ where the excitonic correlation almost vanishes changes from 1.8 for $w=0$ to 2.0 for $w/d=0.64$.

The reason for the increase of the excitonic correlation and shift of the phase boundary as $w$ increases is understood from the behavior of the pseudopotentials.
The increase of inter-layer pseudopotential $V_0^\text{e}$ is favorable for the realization of the excitonic phase, because of the fact that the excitonic phase is the exact ground state for system with $V_0^\text{e}>0$, and all other pseudopotentials being zero.
The decrease of $V_1^\text{a}$ is also favorable for the excitonic phase, since this interaction is responsible for the destruction of the excitonic phase as we will see in the next section.
The effect of finite $w$ through other pseudopotentials should be small, since the deviation from the values at $w=0$ is much smaller for these pseudopotentials compared with $V_0^\text{e}$ and $V_1^\text{a}$

\begin{figure}
\begin{center}
\includegraphics[width=8cm]{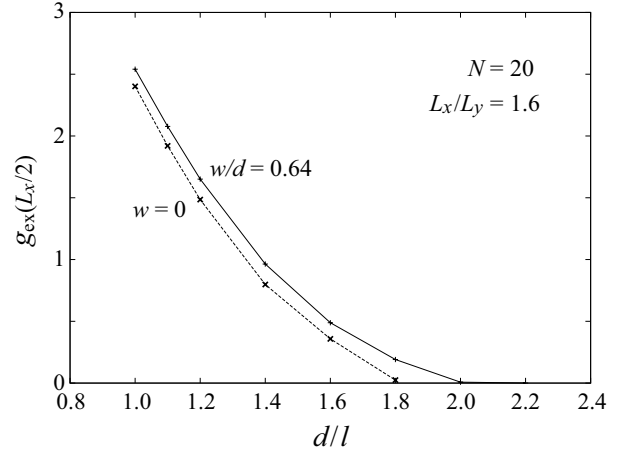}
\end{center}
\caption{
Excitonic correlation for double quantum well of layer thickness $w=0$ and $w/d=0.64$.}
\label{fig:3}
\end{figure}

\subsection{Comparison with experiment}

First, we examine how samples with different values of $w/l$ give different critical values $(d/l)_{\rm c}$.
We compare experiments where the Zeeman splitting is not enhanced artificially.
Thus the transition may be between spin-polarized excitonic phase and partially spin-polarized CF liquid state, but qualitative effect can be seen.
The sample with smallest $w/d$ is used by Wiersma et al.\cite{Wies}.
For this sample with $w/d=0.58$, the critical value $(d/l)_{\rm c}$ is 1.65, and here $w/l=0.96$.
On the other hand, for typical samples with $w/d=0.64$, the critical value $(d/l)_{\rm c}$ is 1.85, and  $w/l=1.18$.
These experimental results qualitatively agree with the present theoretical result.

For more quantitative comparison, we need to compare theoretical result with experiments where the Zeeman splitting is enhanced by in-plane magnetic field such that both phases are completely spin-polarized.
There have been two such experiments.
Giudici et al.\cite{Muraki} measured the activation energy of the quantum Hall state, which is identified as the excitonic phase, at various values of $d/l$.
They found that as the Zeeman splitting is enhanced, the boundary for the quantum Hall state shifts from $(d/l)_{\rm c} \simeq 1.9$ at no enhancement to $2.3$ for sufficiently enhanced Zeeman splitting.
This value of the phase boundary is considerably larger than our result of $(d/l)_{\rm c}=2.0$.

On the other hand, in a more recent experiment, Finck et al.\cite{Finck} investigated the phase boundary using Coulomb drag experiment.
They also find increase of $(d/l)_{\rm c}$ as in-plane magnetic field is increased, but the value of $(d/l)_{\rm c}$ by their definition starts from 1.75 and saturates at around 1.85.
This value is a little lower than our result.
However, Fig.~1 of ref.~\citen{Finck} allows us another way to determine the critical value.
Namely, we can use the value of $d/l$ at which the Hall drag resistance begins to have finite value.
Considering the fact that the Hall drag should be caused by finite excitonic correlation, this definition will be closer with our definition of $(d/l)_{\rm c}$.
Adopting this definition we find that the critical value increases from $(d/l)_{\rm c} \simeq 1.8$ in the perpendicular magnetic field to $2.0$ in the magnetic field tilted by 66$^\circ$.
In this case agreement with our result is quite good.

The samples these authors used have almost identical values of $d$ and $w$.
We do not know the origin of the difference.
It may be due to slight difference in the sample quality and/or geometry, or may be due to the different methods and definitions to determine the transition point.
It has been pointed out that the values of $d$ and $w$ may not be exact.\cite{ZTNC2,Muraki}
Anyway, we suspect that the most important factor making the difference in $(d/l)_{\rm c}$ between the theory and experiment if any is an additional effect caused by tilting of the magnetic field.
In the experiments, the application of the in-plane field makes the total magnetic field tilting about 60 degrees from the direction perpendicular to the 2-d plane.
The single-electron wave function tends to align with the magnetic field line.
Namely, the wave function no longer has the separable form of eq.(\ref{eq:1.1}), and looks as if the center coordinate $X$ shifts in the direction of the in-plane field as $z$ increases within each quantum well.

This deformation of the wave function brings anisotropy in the 2-d plane, and relative angular momentum no longer is a good quantum number.
Investigation of such a case is not considered in this paper, although part of the effect may be able to be taken into account by calculating angle averaged effective pseudopotentials.

\section{$V_0^\text{e}$-$V_1^\text{a}$ Model}

\begin{figure}
\begin{center}
\includegraphics[width=8cm]{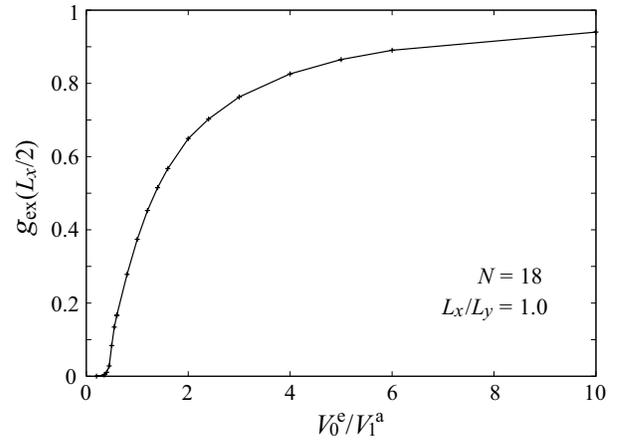}
\end{center}
\caption{
Excitonic correlation for $V_0^\text{e}$-$V_1^\text{a}$ model in torus geometry. 
The total number of electrons $N$ is 18, and the aspect ratio $L_x/L_y$ is 1.
}
\label{fig:4}
\end{figure}

In this section we investigate a minimal model that reproduces phase transition between the excitonic phase and the CF liquid phase. 
The excitonic phase is realized as the exact ground state for a model where only the inter-layer short-range interaction exists between electrons.
Namely, it is a model where only $V_0^\text{e}$ has finite positive value, and other pseudopotentials are zero.
On the other hand, the short-range Coulomb repulsion is essential for the realization of the CF liquid state in a single layer at half-filling.
Namely, a model where only $V_1^\text{a}$ has finite positive value, and other pseudopotentials are zero has ground state where two layers are independent, and in the CF liquid state. 
Therefore, a model in which only $V_0^\text{e}$ and $V_1^\text{a}$ have positive values is the minimal model that has both phases as limiting cases. 
We characterize this model by a single parameter $V_0^\text{e}/V_1^\text{a}$.

The systems we consider in this section are a finite size system on a torus (a rectangular system with periodic boundary conditions) and a system on a sphere. The results of both systems should be the same in the limit of large system. However, in finite systems, size effect appears through the boundary conditions. 
In rectangular systems, the boundary conditions are specified by the area of the unit cell $L_x L_y$ and the aspect ratio $L_x/L_y$, whereas spherical geometry is characterized only by the radius of the sphere $R$. 
We therefore remove the size effect by a simple extrapolation of $R$ in spherical geometry.

The results for the excitonic correlation and charge gap are shown in Figs. 4-7.
In Fig. 4, excitonic correlation for a square system with 18 electrons is plotted as a function of $V_0^\text{e}/V_1^\text{a}$. 
In the limit of $V_0^\text{e}/V_1^\text{a} \rightarrow \infty$, the excitonic correlation $g_{\rm ex}(L_x/2)$ is unity. 
The excitonic correlation decreases with the decrease in  $V_0^\text{e}/V_1^\text{a}$, and almost vanishes at around $V_0^\text{e}/V_1^\text{a} \simeq 0.4$. 
Similar behavior is obtained for other systems with different $N$ and $L_x/L_y$.
The decrease in the excitonic correlation is continuous and the ground state changes to the CF liquid state at some finite value of $V_0^\text{e}/V_1^\text{a}$.

\begin{figure}
\begin{center}
\includegraphics[width=8cm]{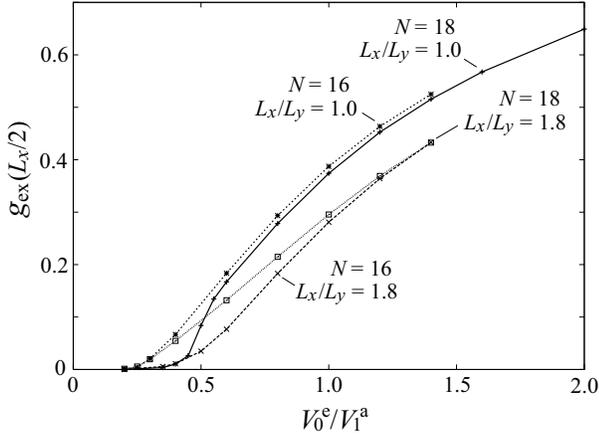}
\end{center}
\caption{
Excitonic correlation for $V_0^\text{e}$-$V_1^\text{a}$ model in torus geometry. $N$ is the total number of electrons.
}
\label{fig:5}
\end{figure}

Figure 5 shows a close-up view of the excitonic correlation near the transition point. 
In this figure, those for several other systems are also plotted to see size dependence. In the region of large $V_0^\text{e}/V_1^\text{a}$, $g_{\rm ex}(L_x/2)$ decreases with the increase in $L_x/L_y$. Although excitonic correlation is independent of the distance between the two electrons in the limit of $V_0^\text{e}/V_1^\text{a} = \infty$,
$g_{\rm ex}(r)$ has $r$ dependence for finite value of $V_0^\text{e}/V_1^\text{a}$ and $g_{\rm ex}(L_x/2)$ decreases with the increase in the distance $L_x/2$. Around the transition point, however, different size dependence appears. 
In this region $V_0^\text{e}/V_1^\text{a}$ is small and the intra-layer interaction, $V_1^\text{a}$, is relatively important.
Since the interaction energy from $V_1^\text{a}$ is sensitive to the spatial configuration of electrons in the unit cell, the ground state energy of finite system depends on boundary conditions $L_x/L_y$.
For the system of 16 electrons, each layer has 8 electrons.
The energy from $V_1^\text{a}$ has a minimum at $L_x/L_y\sim 2$ ($2\times 4$ configuration). 
For 18 electrons, each layer has 9 electrons.
Then a minimum of the energy from $V_1^\text{a}$ appears at $L_x/L_y\sim 1$ ($3\times 3$ configuration). 
Such $L_x/L_y$-dependence enhances or reduces the effect of $V_1^\text{a}$ and the transition (crossover) point from the excitonic phase to the CF liquid phase is modified depending on $N$ and $L_x/L_y$.

\begin{figure}
\begin{center}
\includegraphics[width=8cm]{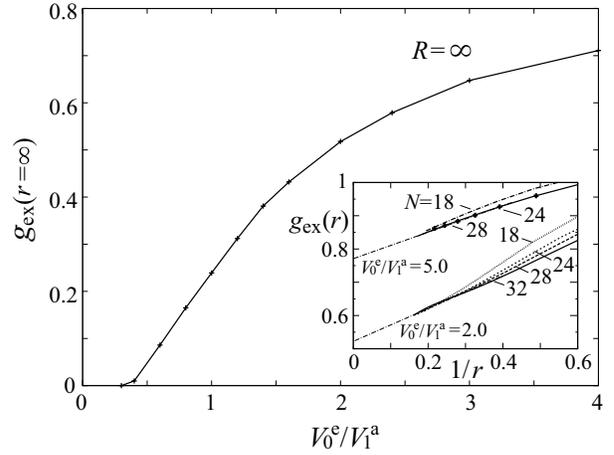}
\end{center}
\caption{
Excitonic correlation for $V_0^\text{e}$-$V_1^\text{a}$ model in spherical geometry. $g_{\rm ex}(r=\infty)$ is the extrapolated value of excitonic correlation in the limit of $r\rightarrow \infty$.
Inset shows $g_{\rm ex}(r)$ for several different $N$. 
Dots in the inset for $N=24$ and $V_0^\text{e}/V_1^\text{a}=5$ represent results in torus geometry of $L_x/L_y=1$.
Other results in the inset are calculated in spherical geometry.}
\label{fig:6}
\end{figure}

To remove such size dependence, we calculate $g_{\rm ex}(r)$ in spherical systems and extrapolate the results to the limit of large distance $r$ and radius $R$ as shown in the inset of Fig.~6.
The extrapolated results are plotted in the main panel of Fig.~6. Here, $g_{\rm ex}(r=\infty)$ is the excitonic correlation between the electrons whose distance $r$ is infinity, that means $g_{\rm ex}(r=\infty)$ represents macroscopic coherence.
Similarly to $g_{\rm ex}(L_x/2)$ of finite systems, macroscopic coherence continuously decreases with the decrease in $V_0^\text{e}/V_1^\text{a}$ and almost vanishes at around $V_0^\text{e}/V_1^\text{a} \simeq 0.4$.

Finally in Fig. 7 charge excitation gap $\Delta_{\text{c}}$ is plotted.
In the limit of $V_0^\text{e} = \infty$, infinite energy is required to add or remove an electron, because at least one exciton consisting electron-hole pair is destroyed. This charge excitation energy is proportional to  $V_0^\text{e}$, and $\Delta_{\text{c}}$ decreases with the decrease in $V_0^\text{e}$. 
At the transition to the CF liquid state,  $\Delta_{\text{c}}$ should vanish in the limit of large system.
In finite systems, however, discrete energy levels leave finite value of charge gap even in the CF liquid state.
In such a case, the charge gap has considerable size dependence and it tends to decrease with the increase in the system size.
The charge gap in Fig. 7 shows large size dependence for small $V_0^\text{e}/V_1^\text{a}$ indicating that the system is in the CF liquid state, but such size dependence is removed for larger $V_0^\text{e}/V_1^\text{a}$.
The change in the behavior of the size effect occurs at around $V_0^\text{e}/V_1^\text{a}\simeq 0.4$ and the gap seems to close in large systems for $V_0^\text{e}/V_1^\text{a} < 0.4$.
We also notice that between $0.4 < V_0^\text{e}/V_1^\text{a} <1.0$ the charge gap deviates from the linear behavior at $1.0 < V_0^\text{e}/V_1^\text{a}$.
This indicates that the excitonic correlation becomes weaker in this regime, and the energy of a charge excitation such as meron pair\cite{MMY} that does not destroy an exciton becomes lower than that of a simple charge excitation that destroys an exciton.
 
These results clearly show that the excitonic correlation characterizing the excitonic phase decreases with the decrease in $V_0^\text{e}/V_1^\text{a}$, and continuously vanishes at around $V_0^\text{e}/V_1^\text{a}\simeq 0.4$ with the transition to the CF liquid state.
This behavior is quite similar to the phase transition (crossover) of the actual system with Coulomb interaction as the separation becomes larger.

\begin{figure}
\begin{center}
\includegraphics[width=8cm]{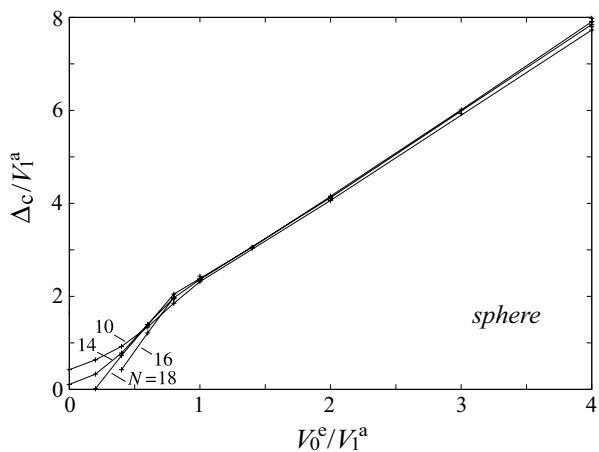}
\end{center}
\caption{
Charge gap for $V_0^\text{e}$-$V_1^\text{a}$ model in spherical geometry. $N$ is the total number of electrons.
}
\label{fig:7}
\end{figure}

\section{Discussion}

We have seen that the excitonic phase stabilized by the inter-layer pseudopotential $V_0^\text{e}$ is destroyed by intra-layer pseudopotential $V_1^\text{a}$.
In the excitonic phase, every single electron state is occupied that is constructed as a superposition of the wave functions of the same center coordinate from two layers:
\be
\Psi = \prod_X \frac{1}{\sqrt{2}}(c_{\sigma,X}^\dagger+\E^{\I\theta} 
c_{-\sigma,X}^\dagger)|0\rangle\,,
\ee
where $\theta$ is an arbitrary phase, and $X$ is the center coordinate in the Landau gauge.
The same wave function can be written in symmetric gauge, in which the angular momentum is a good quantum number.
In this ground state, for an electron in a single electron state of $\sigma$-layer at quantum number $X$, there is no electron in the other layer at the same $X$.
Namely, an electron at $X$ in $\sigma$-layer is always bound to a hole in the $-\sigma$-layer.
Thus, this is a state where the pseudopotential $V_0^\text{e}$ is completely avoided.
On the other hand, an electron has a nearest neighbor electron in the same layer with relative angular momentum 1 with probability 1/2 in this state.
Thus, the pseudopotential $V_1^\text{a}$ makes the ground state energy higher.
In the CF liquid state, this interaction is partly avoided as each electron bounds a hole to form an in-plane dipole.
Namely, if we compare the pair correlation function between the excitonic phase and the CF liquid phase, the CF liquid phase has smaller value at around $r/l=2$ where the pseudopotential $V_1^\text{a}$ has a peak.\cite{Shib2}
The intra-layer pseudopotential $V_1^\text{a}$ acts to move the hole in the opposite layer to the same layer, and weakens the excitonic correlation.
This mechanism of the phase transition is what we suggested in the previous paper,\cite{Shib2} so the essence of the transition is reproduced in the present minimal model.

We have investigated models where $V_3^\text{a}$ and $V_5^\text{a}$ are included also.
We have found that the excitonic phase is stabilized by adding positive $V_3^\text{a}$, but destabilized by adding positive $V_5^\text{a}$.
This behavior is understood considering the mechanism of the phase transition and the shape of the pseudopotentials in the real space.
The two-electron state with relative angular momentum one ($m=1$) has a peak probability for a configuration in which two electrons are separated by distance $r=2l$.
The potential $V_1^\text{a}$ has a peak at this separation, but $V_3^\text{a}$ has a dip here.\cite{Shib4}
Thus, $V_3^\text{a}$ has an effect to weaken the repulsion of $V_1^\text{a}$ at $r=2l$, and reduces the pressure to bound a hole at this distance.
On the other hand, $V_5^\text{a}$ has a peak around $r=2l$, so positive $V_5^\text{a}$ is harmful to the excitonic phase.
These results show importance of the intra-layer interaction at distance $r=2l$ for destruction of the excitonic phase.
The excitonic phase cannot have a hole in the same layer at distance $r=2l$, so the repulsive interaction at this distance destroys the excitonic correlation.

In \S 2, we have seen that the layer thickness stabilizes the excitonic phase to larger value of $d/l$.
This is a reasonable result, since main consequence of the layer thickness is to make $V_0^\text{e}$ larger and to make $V_1^\text{a}$ smaller.

In summary we have shown that the layer thickness has an effect to shift the phase boundary to larger value of $d/l$.
It qualitatively explains the experimental fact that sample with small (large) $w/l$ has smaller (larger) $(d/l)_{\rm c}$.
For typical sample with $w/l=0.64$, quantitative comparison is made.
We partly explained $(d/l)_{\rm c}$ larger than 1.8 observed in recent experiments\cite{Muraki,Finck} in which the Zeeman splitting is enhanced by in-plane magnetic field.
We also investigated a minimal model that show the transition between the excitonic phase and the CF liquid state, and clarified importance of $V_1^\text{a}$ or the repulsive interaction at distance $r\simeq 2l$ for destruction of the excitonic phase.

\section*{Acknowledgment}
The present work is supported by a Grant-in-Aid No. 18684012 from MEXT, Japan.


\begin{thebibliography}{99}
\bibitem{note1} Sample dependence of the critical value $(d/l)_{\rm c}$ is discussed in subsection 2.3.
\bibitem{note2} Equation (4) is simpler than eq.~(5), because for $\sigma=+$ and $\sigma'=-$ (for $V^{\rm e}$) the factor in the integrand of eq.(3), $\E^{-|z_1-z_2|q}$ is equal to $\E^{-(z_1-z_2)q}$, and the $z_1$ and $z_2$ integrations in eq.~(4) are done independently.  It should be noted that this result is valid only for $d>w$, so that simply letting $d\to 0$ in eq.(4) does not give eq.(5).
\end{thebibliography}
\end{document}